\documentstyle[12pt,a4]{article}

\begin{document}
\vspace*{-2cm}
\vspace{1.5cm}
\begin{center}
\baselineskip 5ex
{\LARGE\bf Triplet np Final State Interactions at Large Momentum Transfers}
\\[6ex]
\baselineskip 3ex
Alain Boudard\footnote{Electronic address boudard@phnx7.saclay.cea.fr}\\[1ex]
Service de Physique Nucl\'eaire, CEA Saclay, F-91191 Gif-sur-Yvette, France
\\[4ex]
G\"oran F\"aldt\footnote{Electronic address faldt@tsl.uu.se}\\[1ex]
Division of Nuclear Physics, Uppsala University, Box 535,
S-751 21 Uppsala, Sweden\\[4ex]
Colin Wilkin\footnote{Electronic address cw@hep.ucl.ac.uk}\\[1ex]
University College London, London, WC1E 6BT, UK\\[6ex]
\end{center}

\begin{abstract}
\baselineskip 4ex
Using the simple relation between the $np$ scattering and bound state wave
functions, the spin-triplet contributions to the $pp \to \pi^+(pn)$ and
backward $dp \to p(pn)$ reactions are estimated for low $np$ excitation
energies. The good agreement with the pion production data at 400 and 450~MeV
over a range of angles shows that spin-singlet final states cannot be more than
of the order of 10\%. On the other hand it is seen that the spin-singlet states
must provide about 30\% of the strength in pion production at 1~GeV as well as
deuteron-proton backward scattering at 1.6~GeV.
\end{abstract}
\newpage
\baselineskip 5ex

It is well known that when a scattering wave function is extrapolated to a
bound state pole then the result is proportional to the bound state wave
function. Less appreciated is the fact that the constant of proportionality is
{\it independent} of the form of the potential \cite{FW4,FW5}. Thus for 
single-channel S-wave neutron-proton scattering,
\begin{equation}
\label{1}
\lim_{k\to i\alpha_t}\left\{-\sqrt{\frac{\alpha_t(k^2+\alpha_t^2)}{2\pi}}\:
e^{i\delta}{\psi}_{k}^{t\,(-)}(r)\right\}=\psi_{d}(r)\:.
\end{equation}

\noindent
Here $\alpha_t=\sqrt{m\epsilon}=0.232\:$fm$^{-1}$, where $m$ is the average
nucleon mass, $\epsilon$ the deuteron binding energy, and $\delta$
the uncoupled S-wave triplet phase shift at relative momentum $\vec{k}$.
Though this relation is only true \underline{at} the deuteron pole,
nevertheless for c.m. energies below about 40~MeV and internucleon distances $r$
less than about 1.5~fm the $^3$S$_1$ solution to the Paris neutron-proton 
potential \cite{Paris} empirically behaves like
\begin{equation}
\label{2}
\left|{\psi}_{k}^{t\,(-)}(r)\right|^2\approx 
\frac{2\pi}{\alpha_t(k^2+\alpha_t^2)}
\:\left|{\psi}_{d}(r)\right|^2
\end{equation}
to an accuracy of typically 5\% \cite{FW5}. Thus for large momentum transfer 
reactions such as $\pi$ \cite{FW5} or $\eta$ \cite{FW4} production, which are 
sensitive to the values of the wave functions at short distances, one can get a
simple estimate of the amplitude for the production of a particle $c$, with a
neutron-proton pair in the final state in terms of the amplitude for deuteron
$(d)$ production, {\it viz.}
\begin{equation}
\label{3}
M(ab\to c\{np\}_{k}) \approx - 
\sqrt{\frac{2\pi m}{\alpha_t(k^2+\alpha_t^2)}}\:e^{-i\delta}\:M(ab\to cd)\:.
\end{equation}
      
Though there is no bound state amplitude to set the scale, the momentum
dependence of the S-wave Paris singlet wave function is given by a similar
$\sqrt{k^2+\alpha_s^2}$ factor to that of eq.(1), but with a different pole
parameter $\alpha_s=-0.104\:$fm$^{-1}$ \cite{FW5,Paris}. This factor in turn
determines the energy variation of the production of $np$ singlet final
states. Such a simple model for triplet and singlet production gives a good
description of the total cross sections for $np\to np\eta$ \cite{FW4} and $pp
\to pn\pi^+$ and $pd \to pd\pi^0$ \cite{FW5} near threshold. It is the aim of
the present work to show that the model is equally successful at higher 
energies providing the neutron-proton system is constrained to be at low
excitation energy.

Choosing the same normalisation for bosons and fermions, the c.m. differential
cross section for the production of an N-body final state is
\begin{equation}
\label{4}
d\sigma=(2\pi)^4 \frac{1}{4p_i\sqrt{s}}\:\delta^{(4)}(\tilde{p}_f-
\tilde{p}_i)\:|M|^2\:
\prod_{j=1}^{N}\frac{d^3p_j}{(2\pi)^32E_j}\:\cdot
\end{equation}
Here $\vec{p}_j$ and $E_j$ are the momentum and energy of the final-state
particle $j$ and $\vec{p}_i$ the incident momentum corresponding to a total c.m.
energy $\sqrt{s}$, whereas $\tilde{p}_i$ and $\tilde{p}_f$ are the total 
initial and final four-momentum vectors.

By introducing the ansatz of eq.(\ref{3}) into the phase space and making
kinematic approximations which are very good for high beam energies and low
excitation energies $Q=k^2/m$ in the final $np$ system, the spin-triplet
contribution to the three-body cross section is predicted to be
\begin{equation}
\label{5}
\frac{d^2\sigma}{d\Omega_c dx}(ab\to c\{np\}) = \frac{p(x)}{p(-1)}\:
\frac{\sqrt{x}}{2\pi(1+x)}\:\frac{d\sigma}{d\Omega_c}(ab\to cd)\:.
\end{equation}

The dependence on the energy of particle $c$ is mainly through the explicit
function of $x$, where
\begin{equation}
\label{6}
x= \frac{Q}{\epsilon} =\frac{k^2}{\alpha_t^2}\:\cdot
\end{equation}
$p(x)$ is the c.m.~momentum of particle $c$ in the $c\{np\}$ system, becoming
equal to that in the $cd$ system when $x=-1$. The ratio $p(x)/p(-1)$ is
independent of angle and depends weakly upon $x$ and the incident beam energy.

The most extensive measurement of the $pp\to\pi^+pn$ cross section and
analysing power was carried out several years ago at TRIUMF at 400 and 450~MeV
at several fixed laboratory angles \cite{Falk}. Unfortunately the data are no
longer available in numerical form but it is possible to read values from the
published work \cite{Falk} and the original hand-drawn graphs \cite{Falk2} to 
a fair accuracy. It is known over a wide energy domain from threshold up to the
$\Delta$-region that the final spin-triplet states dominate the reaction
\cite{Sum} and indeed it was noted that the $pp\to \pi^+pn$ analysing powers
follow closely those of $pp\to\pi^+ d$ at the same c.m. pion production angle
\cite{Falk}. 

On the basis of eq.(\ref{5}), the triplet contribution to the cross section
ratio is
\begin{equation}
\label{7}
R = 2\pi\epsilon\:\left.
\frac{d^2\sigma}{d\Omega_c dQ}(ab\to c\{np\})\right/
\frac{d\sigma}{d\Omega_c}(ab\to cd)=
\frac{p(x)}{p(-1)}\,\frac{\sqrt{x}}{1+x}\:.
\end{equation}
A crucial point is that the same $pp\to\pi^+d$ cross sections employed in the
normalisation in the original data analysis has to be used when
constructing the experimental value of the ratio. This information is indeed
available \cite{Falk2}.

In fig.~1 all the data at the two energies and eleven laboratory angles are
plotted against $Q$ and their mutual consistency and the agreement with the
predictions of eq.(\ref{7}) at small $Q$ is impressive. 
The immediate conclusions that one can infer from this comparison are:
\begin{enumerate}
\item
Since any contributions from spin-singlet and P-wave final states to the
semi-inclusive cross sections are incoherent and non-negative, the
experimental points should not lie below the theoretical curves. Violations 
of this bound occur at at large laboratory angles where, due to
kinematic effects, the experimental resolution is poorer.  The isolation of
unbound final neutron-proton states from the deuteron peak, which is not shown
but which is used to normalise the cross section, becomes more difficult under
such conditions.
\item
Taking into account that eq.(2) may be in error by of the order of 5\%, the
possible amount of spin-singlet final $np$ states must be generally less than 
about 10\% of the spin-triplet.
\item
There is evidence from the more reliable small angle points that, above an
excitation energy of 10~MeV, the curves represent a smaller fraction of the
data, as would be expected from the neglect of P and higher waves in the $np$ 
system.
\item 
It has been pointed out \cite{Falk} that the angular distribution of the proton
analysing power follows closely that of $pp\to\pi^+d$ at low excitation energy.
Apart from the larger angle points, the agreement with eq.(\ref{7}) is good 
over a range of laboratory angles, and so this provides evidence for a similar
behaviour for the differential cross section. However the resolution problem
must be clarified to make this more quantitative. Fortunately the approach will
soon be tested in far greater detail through the new high accuracy $pp\to
\pi^+pn$ TRIUMF data currently being analysed \cite{Falk2}.
\end{enumerate}
\indent

Turning now to a second application of the technique, the momentum transfer 
in elastic
deuteron-proton scattering in the backward direction increases very fast with
beam energy. For deuteron laboratory kinetic energies below $T_d\approx
200$~MeV the process is dominated by one-neutron-exchange and the cross
section then depends on the deuteron wave function at high Fermi momenta. 
On the other hand, for $T_d>800$~MeV the process is expected to be driven 
by pionic degrees of freedom and
virtual $\Delta$-excitation \cite{Craigie}. The gap region between these two
models is less well understood. 

The $\vec{d}p\to pd$ cross section and deuteron tensor analysing power $T_{20}$
have been measured at 180$^0$ for $300\leq T_d\leq 2300$~MeV and data are also 
available on the corresponding three-body reaction $\vec{d}p\to p(pn)$ 
at 400 and 1600~MeV \cite{Boudard}. At 1600~MeV the momentum transfer is so
large that, just as for pion production, the reaction should be dominated by
short-range physics. Since the energy resolution in the data is not as good as
for the pion production experiment previously discussed, smearing of the
predictions for the final deuteron and proton-neutron pairs over the
experimental resolution function is required. Using eq.(\ref{5}), the summed
c.m. cross section can be written as
\begin{equation}
\label{8}
d\sigma\,(dp\to p\{pn\}) 
=\frac{p(x)}{p(-1)}\:\left\{\delta(1+x)+\frac{1}{2\pi}\,\frac{\sqrt{x}}{1+x}
\,\Theta(x)\right\}\frac{d\sigma}{d\Omega_p}(dp\to pd)\,
d\Omega_p\,dx\:,
\end{equation}
where the delta-function reflects the two-body $pd$ final state and the term
involving the theta-function the three-body triplet $p(pn)$.

Convoluting this distribution with an experimental resolution function in $Q$,
taken to be Gaussian with $\sigma = 2.35$~MeV, leads to the  spin-triplet
prediction shown in fig.~2a for $T_d = 1600$~MeV. It must be stressed that it
is the area of the elastic deuteron peak which fixes the normalisation of the
triplet break-up contribution through eq.(\ref{8}).

Unlike the low energy pion production data shown in fig.~1, our estimate leaves
some space  for final $np$ singlet states. The shape of such a cross section is
dominated by a similar $\sqrt{x}/(x+\alpha_s^2/\alpha_t^2)$ form but, because
of the  smaller value of $\alpha_s$, it is more peaked in the near-threshold
region. Tolerable agreement with the data is achieved with a singlet
final-state cross section about $0.4\pm 0.05$ of the triplet at large $Q$,
which is essentially consistent with a purely statistical population.

In contrast, at $T_d=400$~MeV it is not possible to get a good fit to the
data shown in fig.~2b even after including singlet $np$ final states. The 
enhancement at small $Q$ is incompatible with the data at larger values since
$np$ P-waves can only increase the cross section there. This breakdown of
the model is not unexpected since at 400~MeV the one-neutron-exchange term
is still the largest contribution and the short-range limit employed here is
of limited value.

The difference in behaviour at the two deuteron beam energies is also seen in
the deuteron tensor analysing power $T_{20}$. Values of this quantity were 
measured for the sum of the first six bins in $np$ excitation energy as well as
for elastic proton-deuteron scattering \cite{Boudard}. In the 1600~MeV region
the difference $\Delta T_{20}$ between these two seems to be energy independent
and consistent with zero. On the other hand both the elastic $T_{20}$ and
$\Delta T_{20}$ vary fast with energy in the 300 -- 600~MeV region, indicating a
dependence on the details of the deuteron wave function. It is clearly not the
high momentum transfer domain where our approach can be used in a naive way.

The most plausible model for large angle elastic $pd$ scattering at high
energies involves the virtual production and absorption of pions \cite{Craigie}
and the striking difference between the singlet/triplet ratio determined from
pion  production at 400/450~MeV and the $dp \to p(pn)$ reaction at an
equivalent proton beam energy of 800~MeV could bring this into question. There
is however evidence from pion absorption on $^3$He that above the $\Delta$
resonance the  singlet/triplet ratio is much larger than at lower energies
\cite{Sum}. This is backed up by the $pp\to \pi^+ pn$ data taken at 1~GeV
\cite{Abaev} and shown in fig.~3. The absolute resolution is here a little
poorer than for the 1.6~GeV $dp\to p(pn)$ experiment \cite{Boudard} but, when
the latter data are degraded slightly by including small admixtures from
neighbouring bins, the shapes of the two data sets are indistinguishable.
Kinematic differences due to the variation of the pion/proton momentum with $Q$
are negligible at such high energies. Also shown are the fitted curves of
fig.~2a with the slightly increased smearing width ($\sigma = 2.60$~MeV). Pion
production at 1~GeV is therefore also consistent with a singlet/triplet ratio
of around $0.4\pm 0.05$.

This steady increase of the singlet/triplet ratio with beam energy at the
$\Delta$-resonance and above is confirmed in the analysis of $p(p,\pi^+)X$
results at 592~MeV and $\theta=0^0$ \cite{Kurt}. A good fit to this data is
achieved with a ratio of $0.25\pm 0.05$.

We have shown that a plausible description can be given for high momentum
transfer reactions leading to unbound $np$ scattering states at low excitation
energy $Q$ in terms of the corresponding cross section for producing a
deuteron. In the limit that the transition operator is of short range our
evaluation is weakly model-dependent. Our principal conclusions are that any
spin-singlet final states in $pp\to \pi^+ pn$ at 400 and 450~MeV must be 10\%
or less of the triplet, though this steadily increases to 20\% and 40\% at 600
and 1000~MeV respectively. This final figure, which is essentially consistent
with a statistical spin factor of 3:1, also reproduces well the 1600~MeV
$\vec{d}p\to p\,pn$ data, giving some support to the idea that this reaction is
driven by virtual pion production. To go further would require the introduction
of an explicit reaction model as well as a fuller evaluation of the final
neutron-proton wave function including the deuteron D-state. Our calculation
has demonstrated that such a refined model should give reasonable answers.  A
systematic experimental study of the $pp \to \pi^+X$ reaction in the forward
direction as a function of energy should show clearly the evolution of the
singlet/triplet ratio and the importance of non-$\Delta$ contributions to the
reaction. The value of such experiments would be enhanced if they could be
carried out with polarised beam and target since, by angular momentum
conservation, in the forward direction an initial longitudinal spin-spin 
correlation is propagated to the final neutron-proton system.\\

Considerable assistance in using the data of ref.\cite{Falk} was given by
W.R.~Falk and this made our work much easier. The values of the nucleon-nucleon
S-wave singlet and triplet scattering state wave functions were kindly
furnished by B.~Loiseau and we have benefited greatly from his advice in their
interpretation. Valuable discussions with V.~Koptev are also gratefully
acknowledged. This work has been made possible by the continued financial
support of the Swedish Royal Academy of Science and one of the authors (CW)
would like to thank them, the The Svedberg Laboratory and the
Kernforschungzentrum J\"ulich for their generous hospitality.

\baselineskip 3ex
\newpage 
\input epsf
\begin{figure}[t]
\begin{center}
\mbox{\epsfxsize=5in \epsfbox{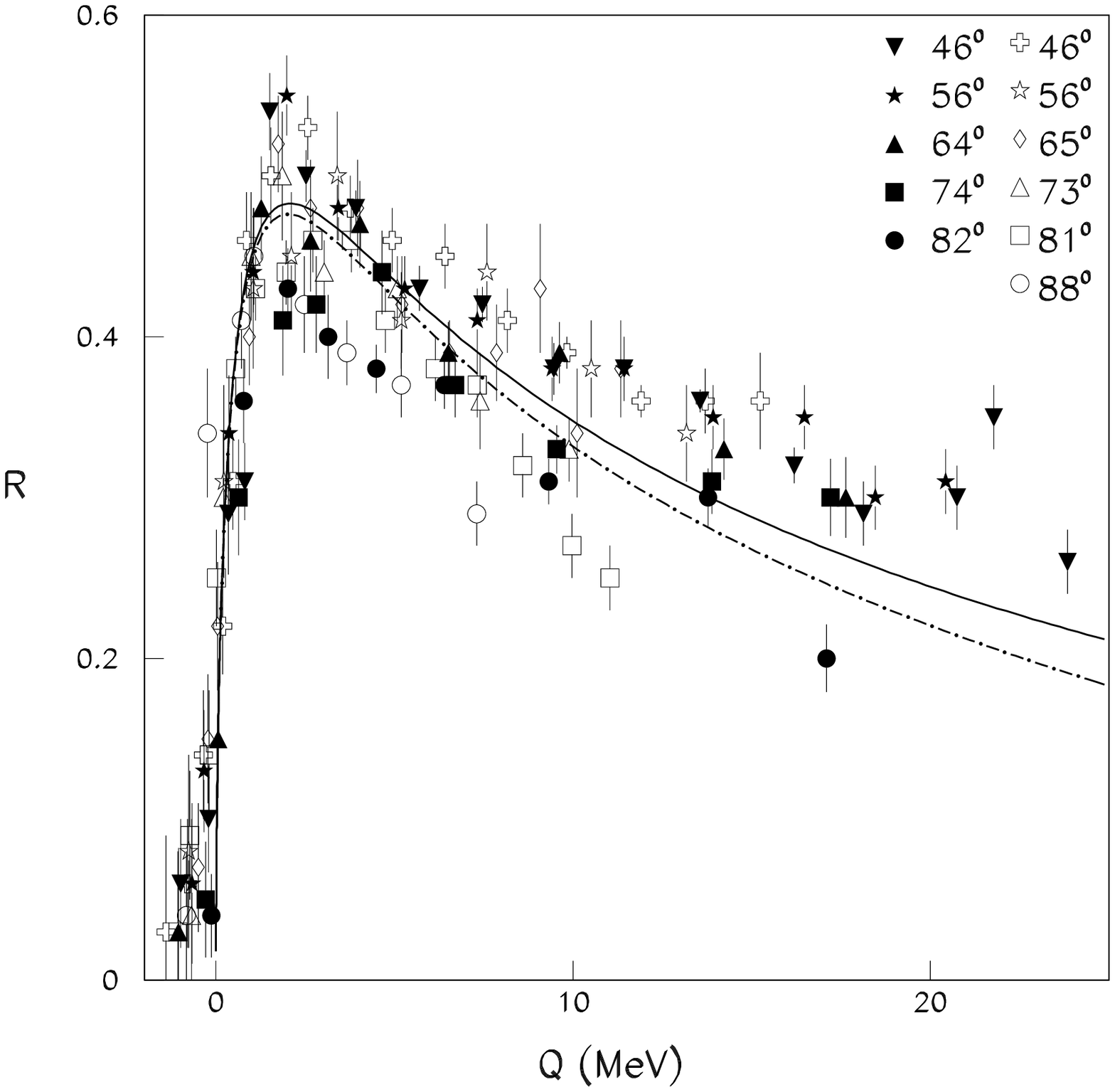}}
\end{center}
\end{figure}

\noindent Figure 1:
The ratio $R$ of the c.m. differential cross section for $pp \to \pi^+ pn$ 
divided by that for $pp\to \pi^+ d$ at the same pion c.m. angle, as defined by
eq.(\ref{7}). The TRIUMF experimental data \cite{Falk} were taken at
400~MeV (open symbols) and 450~MeV (closed symbols) at a total of 11
laboratory angles. The predictions of eq.(\ref{7}) at these two energies
are shown as the broken and solid line respectively, the only difference
between them being due to the change in the variation of the momentum factor.

\newpage
\input epsf
\begin{figure}[t]
\begin{center}
\mbox{\epsfxsize=5in \epsfbox{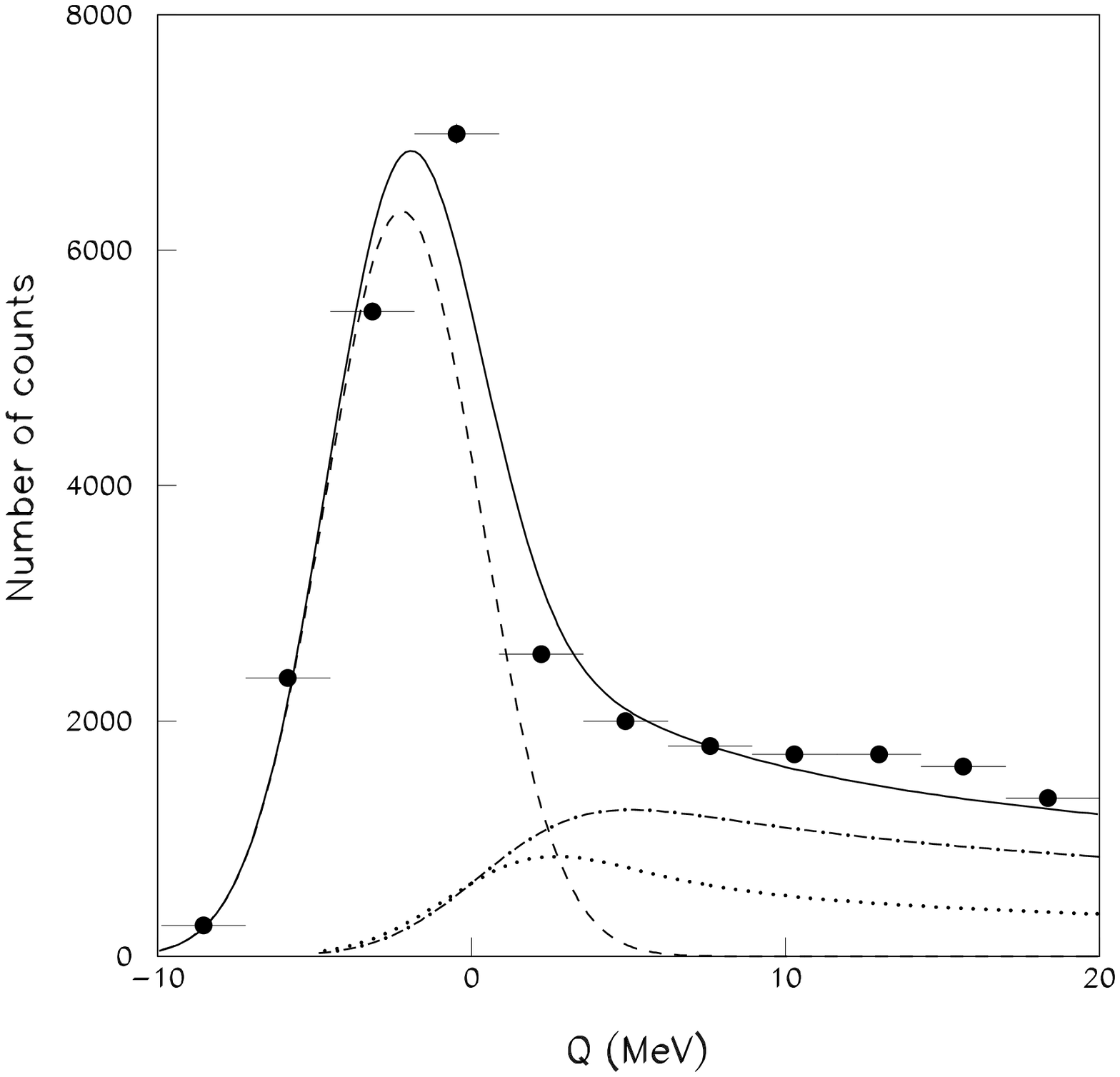}}
\end{center}
\end{figure}

\noindent Figure  2a:
Laboratory differential cross section for the $p(\vec{d},p)pn$
reaction in the backward c.m. direction at $T_d=1600$~MeV as a function of the
excitation energy $Q$ in the unobserved final $np$ system \cite{Boudard}. The
binning is such that 1 channel corresponds to 2.69~MeV. The form of 
eq.(\ref{8}) was smeared over a Gaussian resolution function with $\sigma=
2.35$~MeV and a binning correction applied before being compared to the data.
The broken line is the fitted deuteron elastic peak and the dot-dashed the
resulting prediction for the triplet final $np$ state. The amount of
spin-singlet contribution required to give the overall fit (solid line) is
shown as a dotted line.

\newpage
\input epsf
\begin{figure}[t]
\begin{center}
\mbox{\epsfxsize=5in \epsfbox{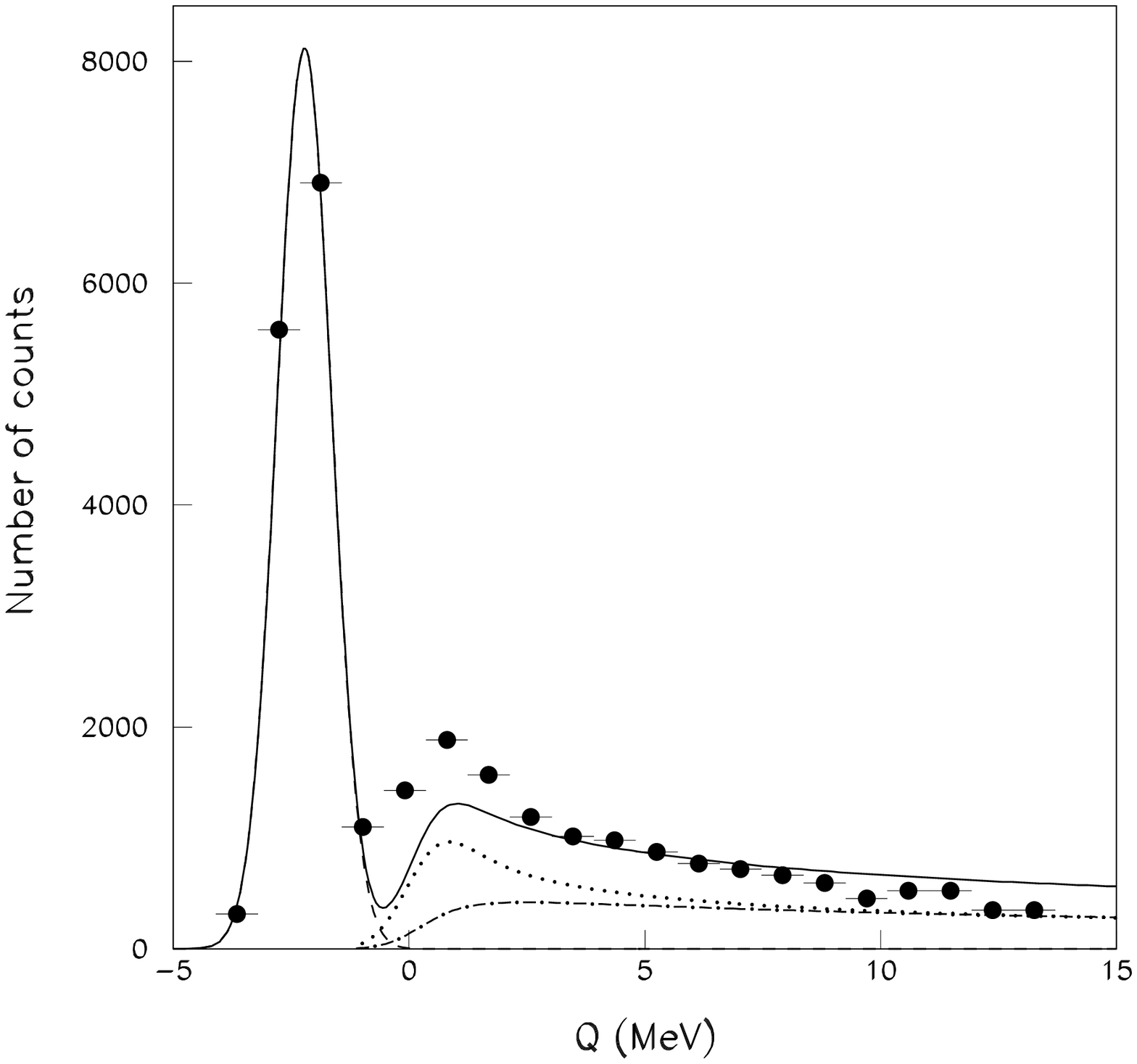}}
\end{center}
\end{figure}

\vspace{1cm}
\noindent Figure 2b:
Laboratory differential cross section for the $p(\vec{d},p)pn$
reaction in the backward c.m. direction at $T_d=400$~MeV as a function of the
excitation energy $Q$ in the unobserved final $np$ system \cite{Boudard}. The
binning is such that 1 channel corresponds to 0.89~MeV. The form of 
eq.(\ref{8}) was smeared over a Gaussian resolution function with $\sigma=
0.52$~MeV and a binning correction applied before being compared to the data.
The broken line is the fitted deuteron elastic peak and the dot-dashed the
resulting prediction for the triplet final $np$ state. 
No satisfactory overall fit to both the low and high excitation energy data 
(solid line) can here be achieved by adjusting the proportion of 
spin-singlet final states (dotted line).

\newpage
\input epsf
\begin{figure}[t]
\begin{center}
\mbox{\epsfxsize=5in \epsfbox{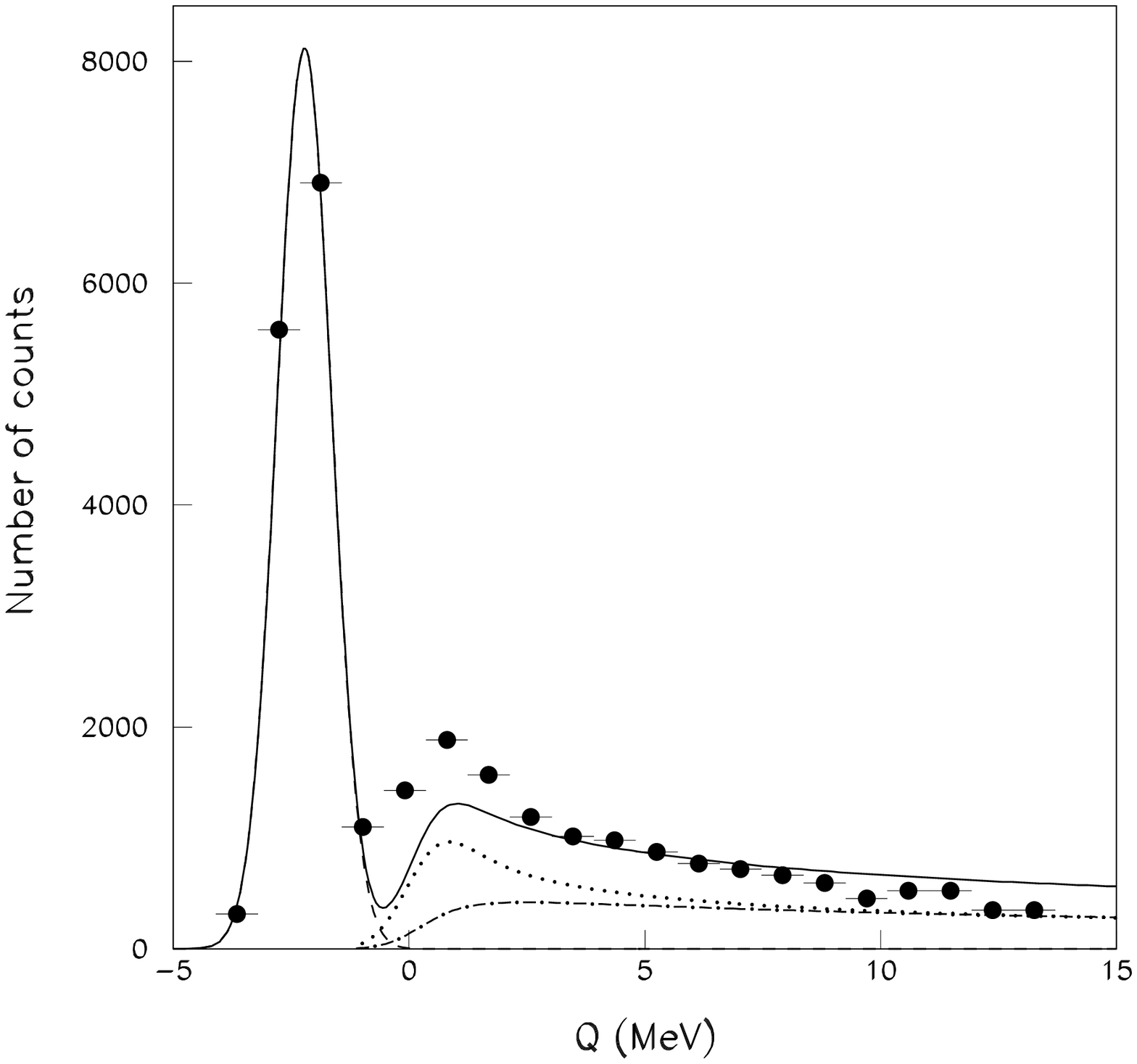}}
\end{center}
\end{figure}

\noindent Figure 3:
Laboratory differential cross section in arbitrary units for the
$p(p,\pi^+)X$ reaction at $\theta = 0^o$ and $T_p=1.0$~GeV (stars) as a
function of the excitation energy $Q$ in the unobserved final $np$ system
\cite{Abaev}. These are compared with the 1.6~GeV $p(d,p)X$ data of
ref.\cite{Boudard} (circles) whose energy resolution has been degraded slightly
by replacing the cross section $\sigma(n)$ in the $n$'th bin by
$0.15\,\sigma(n+1)+0.7\,\sigma(n) +0.15\,\sigma(n-1)$. The fitted curves are
identical to those in fig.~2a but with a smearing parameter of $\sigma =
2.60$~MeV.


\newpage 
\begin{thebibliography}{99} 
%
\bibitem{FW4} G.~F\"aldt and C.~Wilkin, Nucl.Phys. {\bf A604} (1996) 441. 
%
\bibitem{FW5} G.~F\"aldt and C.~Wilkin, Phys.Lett. {\bf B382} (1996) 209. 
%
\bibitem{Paris} M.~Lacombe {\it et al.}, Phys.Rev. {\bf C21} (1980) 861;
M.~Lacombe {\it et al.}, Phys.Lett. {B101} (1981) 139; B.~Loiseau, private
communication (1995). 
%
\bibitem{Falk} W.R.~Falk, E.G.~Auld, G.~Giles, G.~Jones, G.J.~Lolos, 
W.~Ziegler, and P.L.~Walden, Phys.Rev. {\bf C32} (1985) 1972. 
%
\bibitem{Falk2} W.R.~Falk, private communication (1996). 
%
\bibitem{Sum} The experimental situation is well summarised in H.~Hahn 
{\it et al.}, Phys.Rev. {\bf 53} (1996) 1074.
%
\bibitem{Craigie} N.S.~Craigie and C.~Wilkin, Nucl.Phys. {\bf B14} (1969) 477. 
%
\bibitem{Boudard} A.~Boudard, PhD thesis, Universit\'e de Paris-sud (Orsay) 
1983 (unpublished); J.~Arvieux {\it et al.}, Phys.Rev.Lett. {\bf 50} (1983) 19.
%
\bibitem{Abaev} V.V.~Abaev {\it et al.}, Leningrad preprint LNPI-80-569
(unpublished), but see also V.V.~Abaev {\it et al.}, J.Phys. {\bf G14} (1988)
903.
%
\bibitem{Kurt} K.~Gabathuler {\it et al.}, Nucl.Phys. {\bf B40} (1972) 32.
%
\end{thebibliography}
\end{document}